# MMPersistence: A mathematical morphology-oriented software library for computing persistent homology on cubical complexes

Chuan-Shen Hu[a,b]*

[a] *Department of Applied Mathematics, National University of Kaohsiung, 81148 Kaohsiung, Taiwan*

[b] *ORCID: https://orcid.org/0000-0002-4476-7866*

E-mail for the *corresponding author: chuanshenhu.official@gmail.com

Dr. Chuan-Shen Hu received his Ph.D. in Mathematics from National Taiwan Normal University (NTNU), Taiwan, in 2022. He subsequently held postdoctoral fellow positions at the School of Physical and Mathematical Sciences (SPMS), Nanyang Technological University (NTU), Singapore, and at the Department of Mathematics, National Central University (NCU), Taiwan. He is currently an Assistant Professor in the Department of Applied Mathematics at the National University of Kaohsiung, Taiwan. His research focuses on the development of topology- and geometry-based artificial intelligence and computational models for image processing, complex network analysis, and molecular science.

# MMPersistence: A mathematical morphology-oriented software library for computing persistent homology on cubical complexes

Mathematical morphology (MM) is a powerful and widely used framework in image processing. Through set-theoretic and discrete geometric principles, MM operations such as erosion, dilation, opening, and closing effectively manipulate digital images by modifying local structures via structuring elements (SEs), while cubical homology captures global topological features such as connected components and loop structures within images. Building on the GUDHI package for persistent homology (PH) computation on cubical complexes, we propose the *MMPersistence* library, which integrates MM operations with diverse SEs and PH computation to extract multiscale persistence information. By employing SEs of different shapes to construct topological filtrations, the proposed MM-based PH framework encodes both spatial and morphological characteristics of digital images, providing richer local geometric information than conventional cubical homology alone and establishing a unified foundation for analyzing digital images that integrates topological insight with morphological image processing techniques.

## 1. Introduction

Persistent homology (PH) and its visualization through persistence diagrams (PDs), particularly when computed on cubical complexes, play a central role in topological data analysis (TDA) and have been widely applied to the research of digital imaging data (Kaczynski, Mischaikow, and Mrozek 2006). This cubical-complex–based PH framework provides a natural and computationally efficient representation of pixel- and voxel-based data for capturing topological structures within digital images. Applications of PH in imaging span a wide range of fields, including medical and microscopy image analysis, classification, segmentation, and related image processing tasks (Chung and Day 2018; Edwards et al. 2021; Vipond et al. 2021; Aukerman et al. 2022; Pritchard et al. 2023; Chung et al. 2024). Recently, the integration of TDA features from digital images into deep learning architectures has gained increasing attention in image processing tasks, demonstrating the potential of TDA-based representations to enhance model performance as well as improve reliability and interpretability (Hu et al. 2019; Hu et al. 2021; Gupta et al. 2023; Wang et al. 2025).

Regarding a grayscale image as a functional defined on the image domain, with pixel values, typically ranging from 0 to a fixed positive number (e.g., 255), the conventional persistent homology considers the sublevel-set filtration through the thresholding by various pixel values. Concerning the structural intensity encoded by pixel values, this approach focuses on capturing the intensities of the topological structures within the image. In particular, under the sublevel-set filtration, a topological structure represented by a homology class in the induced persistent homology is regarded as more significant and robust if it consists of pixels with extreme intensity values. On the other hand, structures composed of pixels with more moderate intensity values—even when they occupy a larger spatial region within the image—may be regarded as less significant under the sublevel-set filtration, due to their relatively averaged birth-death information (see Figure 1).

Although conventional PH based on the sublevel-set filtration has demonstrated substantial capability in digital image analysis and related applications, its emphasis on topological structures arising from specific ranges of pixel values inevitably neglects certain geometric and spatial characteristics of the underlying image, as illustrated in Figure 1. Figures 1(a) and 1(b) present a grayscale image together with the corresponding PD obtained from the sublevel-set filtration. In contrast, Figure 1(c) shows the same image corrupted by salt noise, where randomly selected pixels are assigned the highest possible intensity, and Figure 1(d) displays the resulting PD. As seen in Figures 1(c) and 1(d), structures associated with extreme pixel values may appear to have disproportionately long lifespans in the PD, even when their true spatial extent within the image is minimal. In particular, when equipped with the bottleneck distance as a metric on (normalized) PDs (Cohen-Steiner, Edelsbrunner, and Harer 2005; Chazal et al. 2016; Bubenik and Elchesen 2022), the PDs shown in Figures 1(b) and 1(d) exhibit a marked discrepancy in their encoded persistence information.

In contrast to a purely topological perspective, *mathematical morphology* (MM) provides a widely used framework for analyzing imaging data that emphasizes geometric and shape-related information within digital images (Serra 1983; Haralick, Sternberg, and Zhuang 1987; Heijmans 1995; Soille 2013; Najman and Talbot 2013). Built upon set-theoretic and discrete structures, MM operations can be computed both feasibly and efficiently. By treating a *structuring element* (SE) as a discrete representation of a local neighborhood around each pixel, MM operations such as *erosion*, *dilation*, *opening*, and *closing* transform a digital image by examining and modifying the pixel values within these neighborhoods. Through this local interaction, MM operations reshape and alter regions of the image according to the chosen SE, thereby offering additional local geometric features for image analysis. These operations constitute a fundamental component of modern image processing (Haralick, Sternberg, and Zhuang 1987; Heijmans 1995). Recently, similar to convolution-based operations in digital image processing, MM operations have been incorporated into deep neural networks, providing an alternative geometric perspective for deep learning architectures in imaging tasks (Franchi, Fehri, and Yao 2020; Shen et al. 2022).

Specifically, Figure 2 provides an overview of the MM-based framework for the geometric and topological analysis of the digital images shown in Figure 1. As illustrated, any given grayscale image can be converted into a binary image through thresholding, and each resulting binary image can then be analyzed using MM-based PH and summarized by its corresponding PD. In contrast to the sublevel-set-based PDs displayed in Figure 1, the average bottleneck distance between the normalized MM-based PDs is markedly smaller. This indicates that MM-based PH is more robust and reliable in the presence of small topological structures with extreme pixel values—features that strongly affect the sublevel-set-based PDs. In this paper, we investigate the underlying reasons for this phenomenon by examining the geometric interpretations of MM-based PDs.

In summary, this paper presents a Python library, called *MMPersistence*, for constructing MM-based filtrations and computing persistent homology for 2D binary digital images. In addition, we demonstrate the functionality of the proposed library on both binary and grayscale images and investigate the geometric interpretation of the birth and death of persistence intervals in the resulting PDs, with particular attention to the spatial information carried by local topological structures in digital images. This work offers an alternative perspective within the TDA community on the application of PH to digital image data.

*1.1. Related work*

Compared with the well-established research and applications of sublevel-set filtrations and their associated persistent homology, mathematical morphology-based filtrations and their persistence represent a comparatively recent direction within the TDA community. The Python library introduced in this paper provides an implementation of MM-based filtrations for digital images and their associated persistent homology computations, drawing on the theoretical foundations developed in our previous studies (Chung, Day, and Hu 2022; Chung et al. 2024).

From a TDA perspective, MM operations—particularly erosion, dilation, opening, and closing—provide a systematic framework for constructing multiparameter filtrations of digital images. In our previous work, we introduced the *alternating sequences of MM operations* to locally manipulate geometric structures within images and thereby induce multiparameter filtrations (Chung, Day, and Hu 2022). This approach offers a complementary viewpoint to existing constructions that are primarily derived from point cloud data. Notably, the resulting persistence diagrams and persistence intervals encode additional spatial information, such as approximations of hole sizes in digital images, which is typically absent from conventional sublevel-set–based filtrations.

For biomedical image analysis, our previous research employed MM-based filtrations and persistent homology, particularly focusing on the opening filtration (see Section 2.3), to analyze mitochondrial structures captured by microscopy images (Chung et al. 2024). This work proposes a morphological bifiltration framework for mitochondrial microscopy images by combining grayscale thresholding with scale-varying morphological opening operations. The resulting bifiltration enables persistent homology to capture both global topological connectivity and local geometric information, such as hole sizes, allowing robust quantitative discrimination between interconnected and fragmented mitochondrial networks.

Recently, MM-based filtrations and persistent homology have attracted increasing attention from the TDA and machine learning communities as effective tools for capturing scale-dependent topological structures in images beyond classical convolutional features and sublevel-set–based persistence. Recent advances demonstrate that MM-induced multiparameter filtrations on cubical complexes admit stable and differentiable vectorizations, enabling their seamless integration into deep learning pipelines (Korkmaz et al. 2025). These developments highlight the potential of incorporating MM-based persistence as a principled topological prior to enhance representation learning and robustness in modern image-based deep learning architectures.

*1.2. Paper organization*

The remainder of the paper is organized as follows. Section 2 introduces the main methodology of MMPersistence, including preliminaries on mathematical morphology (MM), persistent homology (PH) of digital images, and MM-based persistence. Section 3 demonstrates the functionalities of MMPersistence on digital images and interprets the geometric meaning underlying MM-based persistence. Section 4 discusses the limitations of MMPersistence and potential future directions for further investigation and real-world applications. The conclusion of the paper is presented in Section 5.

2. **Materials and methods**

In this section, we present the main theoretical foundations of MMPersistence. Section 2.1 introduces the background and mathematical framework for digital images and mathematical morphology. Section 2.2 briefly reviews homology and persistent homology, together with their geometric interpretations in digital images. Section 2.3 develops the MM-based filtration and its associated persistent homology. Finally, Section 2.4 discusses the MM-based bifiltration for grayscale images and the corresponding persistence.

*2.1. Preliminaries on mathematical morphology*

In this paper, we focus on MM-based persistent homology computation for two-dimensional (2D) digital images defined on a rectangular domain. Notably, we introduce the fundamentals of digital images and MM techniques (Serra 1983; Haralick, Sternberg, and Zhuang 1987; Heijmans 1995; Soille 2013; Najman and Talbot 2013), together with the specific mathematical settings and formulations used in this paper. These preliminaries serve as the foundation for integrating MM methods with tools from TDA.

Mathematically, a *2D rectangle P* is a set of the form $\{a, \ldots, b\} \times \{c, \ldots, d\} \subseteq \mathbb{Z}^2$ for some integers $a \leq b$ and $c \leq d$, and a *2D digital image* is defined as a function $f: P \to \mathbb{R}_{\geq 0}$ that sends each pixel to a nonnegative real number. In practice, the image of function $f$ is $\{0, 1, \ldots, 255\}$ for representing grayscale images and $\{0,1\}$ for binary images. Furthermore, to formalize morphological operations, we denote by $\mathcal{I}_P$ the set of all digital images defined on a 2D rectangle $P$, and by $\mathcal{B}_P$ the set of all binary images defined on $P$. In addition, a natural partial order $\leq$ on $\mathcal{B}_P$ and $\mathcal{I}_P$ is defined by setting $f \leq g$ if and only if $f(x) \leq g(x)$ for all $x \in P$.

A well-known image processing operation called *thresholding* is widely used to transform a grayscale image into a binary one. Specifically, for any given threshold $t \in \mathbb{R}_{\geq 0}$, the thresholding operation is defined as a function $\tau_t: \mathcal{I}_P \to \mathcal{B}_P$ given by $\tau_t(f)(x) = 0$ when $f(x) \leq t$ and $\tau_t(f)(x) = 1$ otherwise. It is straightforward to verify that for any $f, g \in \mathcal{I}_P$, we have $f \leq g$ if and only if $\tau_t(f) \leq \tau_t(g)$ for all thresholds $t \in \mathbb{R}_{\geq 0}$. Furthermore, by the definition above, larger thresholds produce darker binary images; that is, $\tau_t(f) \leq \tau_s(f)$ whenever $s \leq t$.

Associated with a given structuring element (SE)—a subset of the 2D lattice $\mathbb{Z}^2$—the fundamental morphological operations of *erosion*, *dilation*, *opening*, and *closing* can be defined. Specifically, an SE $B$ is a nonempty subset of $\mathbb{Z}^2$ containing the origin $\mathbf{0} = (0,0)$, and the corresponding erosion $\epsilon_B$ and dilation $\delta_B$ operations are defined as functions $\epsilon_B, \delta_B: \mathcal{I}_P \to \mathcal{I}_P$ relative to this SE as follows:

$$\epsilon_B(f)(x) = \min_{\substack{b \in B \\ x+b \in P}} f(x+b) \quad \text{and} \quad \delta_B(f)(x) = \max_{\substack{b \in B \\ x-b \in P}} f(x-b), \quad (1)$$

where $\epsilon_B(f)(x)$ and $\delta_B(f)(x)$ denote the pixel values of images $\epsilon_B(f)$ and $\delta_B(f)$ at the pixel $x$. Then, the opening and closing are functions $O_B, C_B : \mathcal{I}_P \to \mathcal{I}_P$ defined by

$$O_B = \delta_B \circ \epsilon_B \quad \text{and} \quad C_B = \epsilon_B \circ \delta_B, \quad (2)$$

respectively. Changing the SE $B$ may lead to different morphological operations $\epsilon_B, \delta_B, O_B, C_B$ accordingly. Notably, these operations coincide with the identity function on $\mathcal{I}_P$ when $B = \{\mathbf{0}\}$.

Some properties of MM operations with various SEs are useful for constructing topological filtrations (see Sections 2.2 and 2.3). For example, for SEs $B_1$ and $B_2$ satisfying $B_1 \subseteq B_2$, one has

$$\epsilon_{B_2}(f) \leq \epsilon_{B_1}(f) \quad \text{and} \quad \delta_{B_1}(f) \leq \delta_{B_2}(f) \quad (3)$$

for any image $f \in \mathcal{I}_P$. These two ordered relations follow immediately from the definitions in Equation (1). However, for the opening and closing operations, the condition $B_1 \subseteq B_2$ is not sufficient to guarantee that the inequalities $O_{B_2}(f) \leq O_{B_1}(f)$ and $C_{B_1}(f) \leq C_{B_2}(f)$ hold for all $f \in \mathcal{I}_P$ (see, for example, Heijmans 1995; Hu 2022).

In the demonstrations of this paper, we focus on square SEs, denoted by $S_n$ for $n = 1, 2, \ldots$, satisfying $S_1 \subseteq S_2 \subseteq \cdots$. Specifically, the square SEs can be defined by the following recursive formulation: Set $S_1 = \{\mathbf{0}\}$ and define

$$S_n = \begin{cases} S_{n-1} \cup (S_{n-1} + e_1) \cup (S_{n-1} + e_2) \cup (S_{n-1} + e_1 + e_2), & \text{if } n \text{ is even} \\ S_{n-1} \cup (S_{n-1} - e_1) \cup (S_{n-1} - e_2) \cup (S_{n-1} - e_1 - e_2), & \text{if } n \text{ is odd} \end{cases} \quad (4)$$

where $e_1 = (1,0)$, $e_2 = (0,1)$, and $\pm$ denote the Minkowski sum and Minkowski difference of sets in the Euclidean plane. Square SEs have been widely used in constructing MM-based filtrations of digital images (Chung, Day, and Hu 2022; Chung et al. 2024).

Figure 3 illustrates an example of a binary image induced from a grayscale image by thresholding, together with the corresponding images obtained by applying the morphological operations $\epsilon_B, \delta_B, O_B,$ and $C_B$. Intuitively, the erosion $\epsilon_B$ and dilation $\delta_B$ operations shrink and expand the white regions of the binary image, respectively. In contrast, the opening $O_B$ and closing $C_B$ operations remove certain small white and black regions, respectively, while preserving the main white and black structural components of the image.

## *2.2. Persistent homology of digital images*

From a topological perspective, determined by the black pixels, each binary image $f \in \mathcal{B}_P$ can be identified as a topological subspace $\mathcal{X}_f$ embedded in $\mathbb{R}^2$. In this paper, we regard each black pixel of a binary image as a closed unit square in $\mathbb{R}^2$, where any two such squares are non-overlapping; in particular, they may intersect only along their boundaries, either at a shared edge or at a shared vertex, but never in their interiors. Consequently, the set of all black pixels of a binary image $f \in \mathcal{B}_P$ gives rise to a collection of closed 2D squares, together with their edges and vertices, forming a *cubical complex* in $\mathbb{R}^2$ (see, for example, Kaczynski, Mischaikow, and Mrozek 2006). The corresponding topological subspace $\mathcal{X}_f \subseteq \mathbb{R}^2$ is defined as the union of all these closed squares and hence is a compact subset of $\mathbb{R}^2$. Panels (d)–(g) in Figure 4 present examples of binary images defined on a $5 \times 5$ rectangular domain $P \subseteq \mathbb{Z}^2$, along with the corresponding topological spaces depicted in panels (h)–(k).

Homology, an algebraic structure induced from a topological space, captures and expresses the structure of connected components and loops within the subspace

$\mathcal{X}_f \subseteq \mathbb{R}^2$ associated with a given binary image $f \in \mathcal{B}_P$. Formally, the 0th and 1st-*(singular) homology groups* of $\mathcal{X}_f$ with coefficients in the binary field $\mathbb{F}_2 = \mathbb{Z}/2\mathbb{Z}$ are denoted by $H_0(\mathcal{X}_f; \mathbb{F}_2)$ and $H_1(\mathcal{X}_f; \mathbb{F}_2)$, or simply $H_0(\mathcal{X}_f)$ and $H_1(\mathcal{X}_f)$, respectively (see, for example, Kaczynski, Mischaikow, and Mrozek 2006; Greenberg and Harper 2018; Munkres 2018).

As a vector space over $\mathbb{F}_2$, the 0th homology group $H_0(\mathcal{X}_f)$ is the $\mathbb{F}_2$-vector space generated by basis elements corresponding to the connected components formed by the black pixels. On the other hand, the 1st homology group $H_1(\mathcal{X}_f)$ $\mathbb{F}_2$-vector space generated by basis elements corresponding to the loop structures (or holes) enclosed by the black pixels. In particular, the 0th and 1st *Betti numbers* of $\mathcal{X}_f$, denoted $\beta_0(\mathcal{X}_f)$ and $\beta_1(\mathcal{X}_f)$, or simply $\beta_0$ and $\beta_1$, are the dimensions of the vector spaces $H_0(\mathcal{X}_f)$ and $H_2(\mathcal{X}_f)$ (over $\mathbb{F}_2$), respectively, and count the numbers of connected components and loops in the topological space $\mathcal{X}_f$. Panels (a)–(d) in Figure 5 presents examples of topological spaces induced from binary images, together with their corresponding Betti number pairs $(\beta_0, \beta_1)$.

Persistent homology (PH) extends classical homology by considering a filtered sequence of topological spaces, called a filtration, and by tracking the evolution of topological features across the filtration. Mathematically, a filtration of topological spaces is a towered sequence

$$\mathcal{X}_0 \subseteq \mathcal{X}_1 \subseteq \mathcal{X}_2 \subseteq \cdots \subseteq \mathcal{X}_n \tag{5}$$

of topological spaces $\mathcal{X}_i$ such that each $\mathcal{X}_i$ is a subspace of $\mathcal{X}_{i+1}$. Typically, the topological space $\mathcal{X}_0$ is set to be the empty set. By the functoriality of homology, the filtration in Equation (5) induces the following sequence of vector spaces and linear transformations:

$$0 \xrightarrow{\phi_0} H_q(\mathcal{X}_1) \xrightarrow{\phi_1} H_q(\mathcal{X}_2) \xrightarrow{\phi_2} \cdots \xrightarrow{\phi_{n-1}} H_q(\mathcal{X}_n), \quad (6)$$

where $q = 0, 1$ is the dimension that is focused on the persistence computation, and the linear transformations $q = 0, 1$ are induced from the topological inclusions $\mathcal{X}_i \subseteq \mathcal{X}_{i+1}$ for $i = 0, 1, \ldots, n-1$. Panels (a)–(d) in Figure 5 depicts an example of a filtration of topological spaces $\mathcal{X}_{f_1} \subseteq \mathcal{X}_{f_2} \subseteq \mathcal{X}_{f_3} \subseteq \mathcal{X}_{f_4}$, corresponding to a decreasing sequence of binary images $f_1 \geq f_2 \geq f_3 \geq f_4$. Specifically, the binary images $f_1, f_2, f_3, f_4$ in Figure 4, are obtained by thresholding a grayscale image $f$, depicted in Figure 4(a), via the thresholding operations $\tau_i(f)$ for $i = 1, 2, 3, 4$.

A persistence diagram (PD), or persistence barcode (PB), provides a compact summary of persistent homology by recording the birth-death intervals of topological features across a filtration of geometric objects. With deep connections to dynamical systems, Morse theory, and differential geometry (see, for example, Frosini 1992; Barannikov 1994; Robins 1999), PDs capture how topological structures such as connected components and loops evolve as the underlying space varies along the filtration (Carlsson et al. 2005; Cohen-Steiner, Edelsbrunner, and Harer 2005; Ghrist 2008). In general, a feature that appears in a space $\mathcal{X}_b$ at its birth value is considered more robust if it persists over a larger parameter range before disappearing.

Formally, a PD (or PB), is a multiset of pairs $(b, d)$, called persistence intervals (PIs), which encode the birth and death times $b$ and $d$, respectively, of topological features across a filtration of the form given in Equation (5). Figure 5 illustrates a toy example of a filtration induced by binary images and its associated PD and PB representations. In detail, the first row (panels (a)–(d)) forms a filtration of topological spaces embedded in the Euclidean plane $\mathbb{R}^2$, together with the associated PDs and PBs that track the persistence information of the filtration in dimensions 0 and 1. In this example, the 0th and 1st PDs (or PBs) of the filtration are given by the multisets

$$\text{PD}_0 = \{(1,2),(1,\infty)\} \text{ and } \text{PD}_1 = \{(1,4),(2,3),(2,3),(2,4)\}, \quad (7)$$

which encode the birth and death information of connected components and loops across the filtration. More precisely, two connected components are born in the first image, and one of them merges into the other in the second image, corresponding to the PIs (1,2) and (1, ∞). On the other hand, one loop enclosed by black pixels is born in the first image and dies in the fourth image. In addition, three further loop structures appear in the second image: two of them die in the third image, while the remaining one dies in the fourth image.

From an illustrative perspective, PDs and PBs emphasize different aspects of PIs $(b,d)$. In particular, a PD represents each PI as a point in $\mathbb{R}^2$ and primarily highlights the geometric relationships, such as distances, between these points and the diagonal line $\{(x,x) \mid x \in \mathbb{R}\}$ in $\mathbb{R}^2$. In contrast, a PB emphasizes the lifespans $d-b$ of PIs and the multiplicities of identical intervals. For example, in the $\text{PD}_1$ shown in Equation (7), the PI (2,3) appears with multiplicity two, which is represented by two bars in the corresponding PB The bottom row of Figure 5 displays the corresponding PD and PB, illustrating these distinct representational characteristics.

In this paper, as illustrated in Figures 4 and 5, we focus on topological filtrations and the associated PD and PB information arising from the cubical complex representation of two-dimensional binary images. Furthermore, we investigate the geometric interpretation of the birth–death information induced by mathematical morphology (MM)-based filtrations, which are constructed from digital images through MM operations.

## 2.3. MM-based persistent homology of digital images

Leveraging MM operations, one can construct filtrations of digital images. Specifically, the MMPersistence library implements the erosion, dilation, opening, and closing filtrations derived from a given binary image. Mathematically, given a binary image $f \in \mathcal{B}_P$ and a filtered sequence $B_1 \subseteq B_2 \subseteq \cdots \subseteq B_n$ of SEs, the following filtrations of topological spaces are obtained:

$$\mathcal{X}_f \subseteq \mathcal{X}_{\epsilon_{B_1}(f)} \subseteq \cdots \subseteq \mathcal{X}_{\epsilon_{B_n}(f)} \quad \text{and} \quad \mathcal{X}_{\delta_{B_n}(f)} \subseteq \cdots \subseteq \mathcal{X}_{\delta_{B_1}(f)} \subseteq \mathcal{X}_f. \qquad (8)$$

Note that the ordering in the second filtration is reversed compared with the ordering of the SEs, as dilation with a larger SE erodes away more black pixels (see Equation (3)). Notable, by combining the filtrations in Equation (8), one may also consider the combined filtration as follows:

$$\mathcal{X}_{\delta_{B_n}(f)} \subseteq \cdots \subseteq \mathcal{X}_{\delta_{B_1}(f)} \subseteq \mathcal{X}_f \subseteq \mathcal{X}_{\epsilon_{B_1}(f)} \subseteq \cdots \subseteq \mathcal{X}_{\epsilon_{B_n}(f)}. \qquad (9)$$

It is worth noting that, since $\epsilon_B(f) = f = \delta_B(f)$ for every $f \in \mathcal{I}_P$ when $B = \{\mathbf{0}\}$, the SE $B_1$ in Equations (8) and (9) is typically chosen so that $B \neq \{\mathbf{0}\}$. In particular, this paper primarily considers the square SEs defined in the referenced equation to modify a given digital image and to generate filtrations for persistent homology computations.

A similar filtration can be constructed using the opening and closing operations. Formally, suppose $f \in \mathcal{B}_P$ is a binary image and $B_1 \subseteq B_2 \subseteq \cdots \subseteq B_n$ is a sequence of SEs satisfying

$$O_{B_{i+1}}(f) \leq O_{B_i}(f) \quad \text{and} \quad C_{B_i}(f) \leq C_{B_{i+1}}(f) \qquad (10)$$

for every $i = 1, 2, \ldots, n-1$. Then the following filtrations are obtained:

$$\mathcal{X}_{C_{B_n}(f)} \subseteq \cdots \subseteq \mathcal{X}_{C_{B_1}(f)} \subseteq \mathcal{X}_f \subseteq \mathcal{X}_{O_{B_1}(f)} \subseteq \cdots \subseteq \mathcal{X}_{O_{B_n}(f)}, \qquad (11)$$

and hence the associated PH, together with its PDs or PBs, can be computed. Notably, one can show that for any digital image defined on a rectangular image domain and any filtered sequence of rectangular SEs $B_1 \subseteq B_2 \subseteq \cdots \subseteq B_n$, the inequalities in Equation (10) hold for every (grayscale) image on the image domain (see, for example, Hu 2022). In particular, the filtered sequence of square SEs defined in Equation (4) satisfies this assumption, providing computationally practical filtrations for persistent homology.

More generally, suppose $\gamma_1, \gamma_2, \ldots, \gamma_n: \mathcal{B}_P \to \mathcal{B}_P$ are image operations on binary images such that $\gamma_1(f) \leq \gamma_2(f) \leq \cdots \leq \gamma_n(f)$ for every $f \in \mathcal{B}_P$. Then the following filtration of topological spaces is induced:

$$\mathcal{X}_{\gamma_n(f)} \subseteq \cdots \subseteq \mathcal{X}_{\gamma_1(f)}. \tag{12}$$

The MMPersistence library also provides a platform for computing persistent homology on sequences of binary images whose associated topological spaces satisfy the properties given in Equation (12).

### 2.4. MM-based persistence of grayscale images

Theoretically, the MM-based persistence framework can also be extended to grayscale images. Formally, for any grayscale image $f \in \mathcal{I}_P$, thresholds $0 \leq t_1 < t_2$, and SEs $B_1 \subseteq B_2$ satisfying the conditions in Equation (10), the following partial-order diagrams hold for erosion and opening operations:

$$\begin{array}{ccccccc} \epsilon_{B_1}(\tau_{t_2}(f)) & \to & \epsilon_{B_1}(\tau_{t_1}(f)) & & O_{B_1}(\tau_{t_2}(f)) & \to & O_{B_1}(\tau_{t_1}(f)) \\ \uparrow & & \uparrow & \text{and} & \uparrow & & \uparrow \\ \epsilon_{B_2}(\tau_{t_2}(f)) & \to & \epsilon_{B_2}(\tau_{t_1}(f)) & & O_{B_2}(\tau_{t_2}(f)) & \to & O_{B_2}(\tau_{t_1}(f)) \end{array}, \tag{13}$$

where an arrow $f \to g$ for binary images $f, g \in \mathcal{B}_P$ denotes the relation $f \leq g$. On the other hand, when considering morphological dilation and closing, the diagrams

$$\begin{array}{ccccccc}
\delta_{B_1}(\tau_{t_2}(f)) & \to & \delta_{B_1}(\tau_{t_1}(f)) & & C_{B_1}(\tau_{t_2}(f)) & \to & C_{B_1}(\tau_{t_1}(f)) \\
\downarrow & & \downarrow & \text{and} & \downarrow & & \downarrow \\
\delta_{B_2}(\tau_{t_2}(f)) & \to & \delta_{B_2}(\tau_{t_1}(f)) & & C_{B_2}(\tau_{t_2}(f)) & \to & C_{B_2}(\tau_{t_1}(f))
\end{array} \quad (14)$$

hold, but with the direction of the vertical partial orders reversed. Consequently, by considering the sets of black pixels associated with the images in Equations (13) and (14), one obtains the induced *bifiltrations* of topological spaces.

*Bifiltrations*, or more generally *multifiltrations*, consider filtered topological spaces parameterized by more than one parameter, such as the threshold value and the index of a chosen sequence of SEs introduced above. This MM-based framework provides richer geometric and topological information than one-parameter filtrations and their associated persistent homology, thereby offering more subtle insights for the analysis of imaging data (Chung, Day, and Hu 2022; Chung et al. 2024; Korkmaz et al. 2025).

## 3. Result

The results section of this paper is organized into three parts. Section 3.1 introduces the functionality of MMPersistence when applied to binary images, illustrating the geometric meanings of birth and death values arising from MM-based filtrations and the resulting persistence diagrams. Building on the framework outlined in Section 2.4, Section 3.2 extends this discussion to grayscale images and provides corresponding geometric interpretations. Finally, Section 3.3 describes the programming environment and software dependencies required for running MMPersistence.

### *3.1. MM-based persistence for binary images*

Section 3.1 introduces the persistence of morphological filtrations constructed on binary images, focusing on erosion, opening, dilation, and closing operations, and discusses the computation of PDs as well as the geometric interpretation of the resulting PIs. Notably, this paper primarily focuses on the 1st PDs associated with MM-based filtrations. The geometric analysis of the 0th PDs can be treated as a dual case, obtained by exchanging the roles of white and black pixels in the binary image.

*3.1.1: Persistence based on erosion and opening filtrations*

Erosion and opening operations share the property of progressively enlarging the set of black pixels in a given binary image. Specifically, for any nested sequence of SEs $B_1 \subseteq B_2 \subseteq \cdots \subseteq B_n$, an erosion-based filtration is induced as described in Equation (3), in which the corresponding black regions form an increasing sequence i.e., $\mathcal{X}_f \subseteq \mathcal{X}_{\epsilon_{B_1}(f)} \subseteq \cdots \subseteq \mathcal{X}_{\epsilon_{B_n}(f)}$. Similarly, for sequences of SEs satisfying Equation (10), such as the SEs defined in Equation (4), an opening-based filtration is obtained, where the black pixel sets grow monotonically under successive opening operations i.e., $\mathcal{X}_f \subseteq \mathcal{X}_{O_{B_1}(f)} \subseteq \cdots \subseteq \mathcal{X}_{O_{B_n}(f)}$.

Figure 6 illustrates a representative example of a $920 \times 920$ binary image $f$ containing six 1-dimensional holes, together with the corresponding 0th and 1st PDs associated with the erosion filtration $\mathcal{X}_f \subseteq \mathcal{X}_{\epsilon_{S_1}(f)} \subseteq \cdots \subseteq \mathcal{X}_{\epsilon_{S_{61}}(f)}$ with SEs $S_n$ defined in Equation (4). In this example, holes have rectangular shapes of different sizes, which are reflected in the 1st PD

$$\text{PD}_1 = \{(0, 8), (0, 14), (0, 22), (0, 28), (0, 41), (0, 61)\}. \tag{15}$$

In particular, the death values encode the widths of the rectangular hole structures. Such geometric information about hole sizes is typically omitted in conventional homology computations, rendering these holes topologically indistinguishable. Note that, in this example, the opening filtration induces the same PDs since $O_{S_n}(f) = \epsilon_{S_n}(f)$ for all $n =$

$2, \ldots, 61$.

In contrast to the rectangular holes considered previously, Figure 7 presents an example of a $500 \times 500$ binary image and the corresponding PDs induced by erosion- and opening-based filtrations constructed using the square structuring elements defined in Equation (4) ($n = 2, \ldots, 26$). The input binary image in Figure 7 contains five relatively irregular white curves of varying thickness, which collectively form five hole structures. By applying the erosion and opening filtrations, the associated PDs can be computed. Owing to the geometric irregularity of the holes in the original image, the resulting first PD ($PD_1$) is considerably more complex than that in Equation (15); in particular, it contains a large number of PIs with positive birth values. This phenomenon arises because the erosion process partially removes irregular regions of the curves, thereby generating additional hole structures at filtration values greater than zero. Nevertheless, the PIs with birth value zero correspond one-to-one to the five original holes, with their death values reflecting the thicknesses of the curves. This provides a quantitative and mathematically interpretable description of the size and "thickness" of topological structures in digital images. Furthermore, as illustrated in Figure 7, the PIs with positive birth values form five distinct clusters in $PD_1$, with the overlapping interval $(0,1)$ constituting one of these clusters. In this setting, PIs with similar death values tend to aggregate, giving rise to cluster structures in the PD.

In addition, although both erosion and opening operations remove or weaken white objects in a binary image, the resulting PDs exhibit distinct distributions of PIs. As illustrated in Figure 7, the opening operation tends to eliminate small or thin white structures while preserving more robust ones. In contrast, erosion removes or weakens all white structures uniformly. As a consequence of these differing geometric properties, the $PD_1$ induced by the erosion filtration typically exhibits a more dispersed distribution of PIs, reflecting the presence of numerous fragmented white components.

*3.1.2: Persistence based on dilation and closing filtrations*

Rather than employing erosion- and opening-based operations that detect the sizes of holes embedded in black regions of a binary image, dilation- and closing-based operations promote the connectivity of white regions (i.e., holes). Thin black interfaces separating adjacent holes are more readily eliminated by these operations, leading to the merging of nearby white structures, which is captured by the corresponding death values in the resulting PDs.

Figure 8 presents two $500 \times 500$ binary images and the corresponding PDs and PBs induced by dilation filtrations constructed using the SEs $S_n$ defined in Equation (4) ($n = 2, \ldots, 26$). In these examples, both binary images contain hole structures, but their geometric arrangements differ substantially. Specifically, the holes in the first image are more densely clustered than those in the second image, whereas the holes in the second image are organized into two primary clusters. This geometric distinction is clearly reflected in the induced first PDs. In particular, the $PD_1$ of the second image contains a PI (2,26) with a lifespan of 24, which is significantly longer than any PI observed in the $PD_1$ of the first image. This behavior provides a quantitative and mathematically principled means of characterizing the spatial sparseness and connectivity of hole structures within digital images. In our previous work, dilation- and closing-based filtrations and their associated persistence diagrams were employed to quantify the connectivity of mitochondrial structures in microscopy images, thereby characterizing the robustness of mitochondrial networks (Chung et al., 2024).

Figure 9 presents an additional example of a dilation-based filtration and the associated persistence diagrams. The binary image contains two holes embedded within a black disk region. Although the two holes have identical geometric shapes and sizes, they give rise to different persistence intervals due to their distinct spatial locations within the disk. Specifically, the hole on the left-hand side is closer to the disk boundary than the one on the right-hand side and consequently exhibits a shorter lifespan in the corresponding persistence diagram.

In summary, erosion- and opening-based operations and their associated persistence representations capture not only the size and shape information of hole structures but also their relative spatial locations within a binary image, offering a more informative geometric characterization of topological features than that provided by conventional sublevel-set–based persistence.

### *3.2. MM-based persistence for grayscale images*

Inspired by the observations in Equations (13) and (14), the proposed MM-based PH framework can be extended to grayscale images. Formally, let $f: P \to \mathbb{R}_{\geq 0}$ be a grayscale image and let $t \in \mathbb{R}_{\geq 0}$ be a threshold. A corresponding binary image $\tau_t(f): P \to \{0, 1\}$ is obtained via thresholding. For any MM-based filtration constructed on $\tau_t(f)$, the associated persistence can then be computed. For $q \in \{0,1\}$, we denote the $q$th persistence diagram of a given MM-based filtration of $\tau_t(f)$ by

$$\text{PD}_q^{\zeta \cdot}(f, t), \tag{16}$$

where $\zeta\cdot$ denotes a family of MM operations (e.g., erosion, dilation, opening, and closing operations) on $\mathcal{J}_P$. By considering a collection of thresholds $t_1, t_2, \ldots, t_m$, the following sequence of PDs is obtained:

$$\text{PD}_q^{\zeta \cdot}(f, t_1), \text{PD}_q^{\zeta \cdot}(f, t_2), \ldots, \text{PD}_q^{\zeta \cdot}(f, t_m). \tag{17}$$

Consequently, by aggregating the persistence information shown in Equation (17), such as via direct concatenation, the resulting sequence of PDs provides a multiscale feature representation for an input grayscale image.

Rather than computing the bifiltration and its persistence as defined by the relations in Equations (13) and (14), this work focuses on a feature-representation perspective and presents a straightforward approach for applying MMPersistence to grayscale image analysis. Combining MM-based bifiltrations with multipersistence computational tools for imaging data, such as RIVET (The RIVET Developers 2020), represents a promising direction for capturing multiscale geometric structures and complex spatial interactions.

To quantify the difference between PDs induced by conventional sublevel-set filtrations and those induced by MM-based filtrations, we employ the bottleneck distance (Cohen-Steiner, Edelsbrunner, and Harer 2005; Chazal et al. 2016; Bubenik and Elchesen 2022). Equipped with this distance, the collection of all finite PDs forms a metric space (see, for example, Oudot 2015). In this setting, two PDs with more similar distributions and patterns of PIs have a smaller bottleneck distance, and vice versa. In this paper, in comparison with conventional persistent homology computed from grayscale images, we use the bottleneck distance to demonstrate the advantages of MM-based persistent homology in capturing the shape and geometric information of topological structures within digital images.

To compare PDs whose PIs span different scales, such as those arising from conventional sublevel-set filtrations and MM-based filtrations, we normalize the PDs by rescaling all birth and death values to lie in the interval [0,1]. In detail, a PD derived from a sublevel-set filtration of a grayscale image typically has PIs whose birth and death values lie in the range [0,255]. In contrast, for an MM-based PD constructed from a filtration induced by a sequence of SEs $B_1 \subseteq B_2 \subseteq \cdots \subseteq B_n$ of length $n$, the birth and death values are bounded by $[0, n+1]$, where the value $n+1$ corresponds to the totally black image.

Figure 10 illustrates the computational pipeline of MMPersistence for grayscale images, using the same settings as those in Figures 1 and 2. As shown in the figure, the bottleneck distance between the normalized conventional PDs (or PBs) obtained from the sublevel-set filtrations of the original image $f$ and the noisy image $g$ with added salt noise is

$$d_B(\text{PD}_1(f), \text{PD}_1(g)) \approx 0.4353. \tag{18}$$

In contrast, by fixing a threshold (e.g., $t = 100$) and employing the opening-based filtration to estimate the size information of the salt noise, the bottleneck distance between the corresponding normalized MM-based PDs is

$$d_B(\text{PD}_1^{O_{B.}}(f, 100), \text{PD}_1^{O_{B.}}(g, 100)) \approx 0.091. \tag{19}$$

Figure 2 presents an overview of the results obtained by considering multiple thresholds and the corresponding average bottleneck distances for $t = 50, 100$, and $150$. Notably, the average distance coincides with the value reported in Equation (19), since the three individual distances are nearly identical. This phenomenon arises because most salt-noise artifacts correspond to persistence intervals located near the diagonal, specifically those of the form $(0,0.091)$.

Compared with the conventional PD or PB information solely based on pixel values, MM-based persistence is more sensitive to the size and shape of topological structures in images. This enhanced geometric awareness enables a more informative and robust characterization of image structures, particularly in the presence of noise or complex spatial patterns.

*3.3. Software and code environment*

The MMPersistence library is implemented in Python and is built upon the cubical complex functionality provided by the GUDHI library (Dłotko 2025; The GUDHI Project 2025). The development of MMPersistence and all experiments were conducted on the Google Colab platform using Python version 3.12.12, compiled on Oct 10, 2025, at 08:52:57, with GCC version 11.4.0. Notably, the visualization and plotting of PDs in MMPersistence rely on the Persim package, which is part of the Scikit-TDA project (Saul and Tralie 2019). On the other hand, the GUDHI library supports the visualization of PBs. Further information, including our Python implementation and its dependencies (NumPy, Numba, OpenCV, Matplotlib, etc.), is available in the code availability statement and the linked GitHub repository.

**4. Discussion and future work**

In this section, we discuss the limitations, open questions, and future directions of the proposed MMPersistence library, with particular emphasis on its deeper geometric interpretations and its potential applications to real-world imaging tasks.

*4.1. Limitations*

Although MM-based persistence is able to capture spatial and size-related information of topological structures in digital images, a more rigorous mathematical formulation and theoretical analysis remain to be developed. For example, while dilation- and closing-based operations can connect hole structures, the precise geometric interpretation of persistence intervals in the 1st PD, such as whether they reflect sparseness, connectivity, or locational information as illustrated in Figures 8 and 9, is still not fully understood. In addition, the relationship between persistence intervals in the 0th and 1st PDs remains an open problem, as evidenced by the examples in Figure 8. Furthermore, since erosion, dilation, opening, and closing operations act globally on all pixels of an image, it is challenging to precisely localize the geometric meaning of individual persistence intervals in MM-based PDs, such as identifying the specific hole structures or spatial regions to which they correspond.

On the other hand, beyond a more detailed investigation of the geometric meaning of MM-based persistence for digital images, several computational aspects of MMPersistence warrant further exploration. In particular, improving the efficiency of morphological operations—especially erosion and dilation—would significantly accelerate the construction of MM-based filtrations and the computation of the associated PDs. Moreover, leveraging GPU-based parallelization and optimized low-level implementations represents a promising direction for scaling the framework to larger images and more computationally demanding applications.

## 4.2. Persistence from other Image operations

One direction for future work is to integrate additional image operations, such as convolution, into the construction of topological filtrations for PH computations on digital images, enabling the extraction of richer geometric information for more refined analyses of image structures. In particular, to better capture local topological and geometric information, the development of localized image operations, such as *local erosion*, *local dilation*, *local opening*, and *local closing*, together with their associated filtrations and persistence representations, represents an important and promising approach.

*4.3. Future directions on real-world applications*

In addition to adopting more image processing techniques for filtration construction, future work will extend MM-based filtrations and their PD representations to a broader range of real-world imaging applications. By integrating MM-based persistence to jointly encode topological evolution and morphologically meaningful geometric information in digital images, such as connectivity, fragmentation, and characteristic hole sizes, this framework provides a principled approach for analyzing complex structural patterns beyond what is captured by conventional sublevel-set-based persistence. Promising directions include systematic validation on biomedical and microscopy imaging tasks, such as structural characterization, classification, and quality assessment, building on prior results showing that MM-based filtrations can effectively quantify connectivity and robustness in mitochondrial networks (Chung et al., 2024). More generally, integrating MM-based persistence into machine learning models is a promising direction, either by using vectorized persistence summaries as geometry-aware features or by embedding them directly into learning architectures as topology-informed components.

## 5. Conclusion

In conclusion, this paper investigates the geometric and topological meanings of the substructures arising in topological filtrations built from mathematical morphology techniques on 2D digital images. We propose the Python library MMPersistence and demonstrate its functionality on digital image examples. The results show that this framework provides additional geometric information beyond that captured by conventional persistent homology based on the sublevel-set filtration. Furthermore, by releasing MMPersistence as an open-source resource, this work offers a freely accessible tool within the Python environment, supporting integration with deep neural networks and enabling broader applications in image processing and related real-world tasks.

**Code availability** The Python code for the proposed MMPersistence library is available in the following GitHub repository: https://github.com/ChuanShenHu/MMPersistence. The demonstration images for MMPersistence are also available in the GitHub repository. All images were produced by the author, and no copyright issues are involved.

**Acknowledgements** The major part of this work was initiated and completed during the author's appointment as a postdoctoral researcher in the Department of Mathematics at National Central University (NCU) in Taiwan. Parts of the ideas and theoretical foundations of this work were developed by the author's research during his PhD program in the Department of Mathematics at National Taiwan Normal University, Taiwan. The implementation, geometric interpretation, and demonstration of the proposed code library (MMPersistence) were carried out entirely during the author's employment at NCU. The author gratefully acknowledges financial support from the National Science and Technology Council (NSTC) of Taiwan under Grant No. NSTC 114-2811-M-008-069, hosted by Dr. John M. Hong (NCU). The author used Grammarly and ChatGPT-5 for language refinement; all intellectual content is solely the author's.

**Nomenclature**

| | |
|---|---|
| 2D | 2-dimensional |
| PD | Persistence diagram |
| PB | Persistence barcode |
| PH | Persistent homology |
| PI | Persistence interval |
| MM | Mathematical morphology |
| MMPersistence | The proposed Python library (MM-based persistence) |
| SE | Structuring element |

| TDA | Topological data analysis |

Figure 1. Grayscale images and the persistence diagrams (PDs) obtained via sublevel-set filtration according to pixel values. (a) A grayscale image with pixel values ranging from 0 to 255. (b) The PD representation of the image in (a), containing the 0th and 1st PDs. (c) A modified version of the image in (a) with random salt noise, where the number of salt pixels is given by the floor number of $0.01 \times \text{Height} \times \text{Width}$, with Height and Width denoting the height and width of the image. (d) The PD representation of the image in (c), containing the 0th and 1st PDs. The bottleneck distance between the induced normalized 1st PDs (see Section 2.2) is approximately 0.4353.

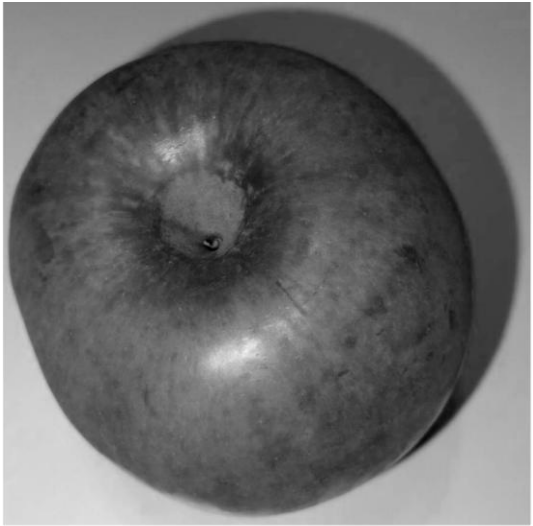

(a)

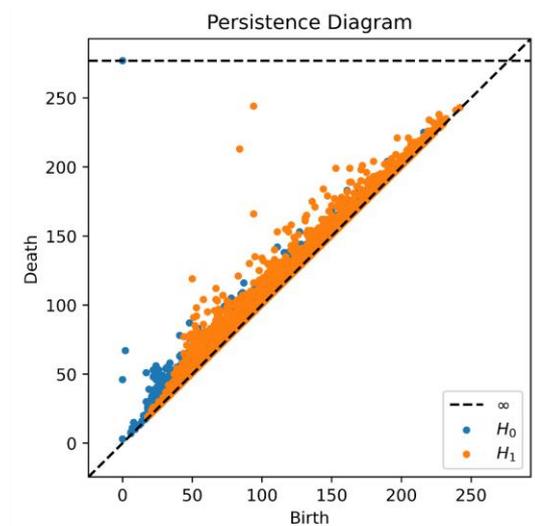

(b)

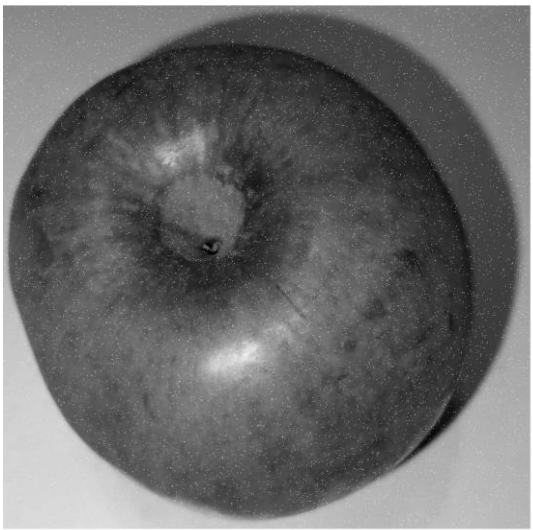

(c)

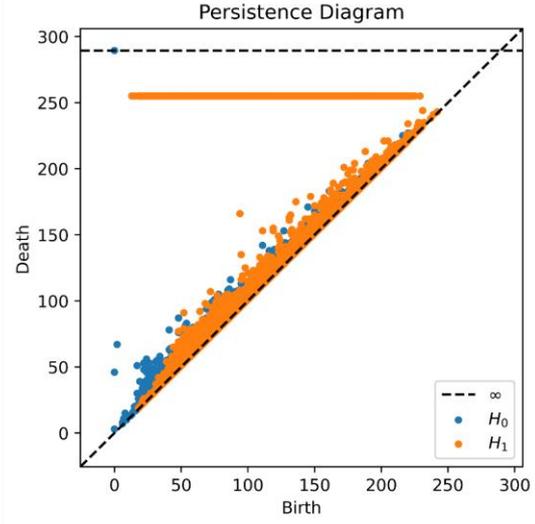

(d)

Figure 2. An overview of MM-based persistence for analyzing grayscale images. The first panel displays the original and noisy grayscale images. The second panel illustrates three pairs of MM-based persistence diagrams (PDs) of thresholded images, computed from the grayscale data using the opening filtration built by square SEs (see Section 2.3). The third panel presents the average bottleneck distance between the induced normalized 1st MM-based PDs, which is 0.091.

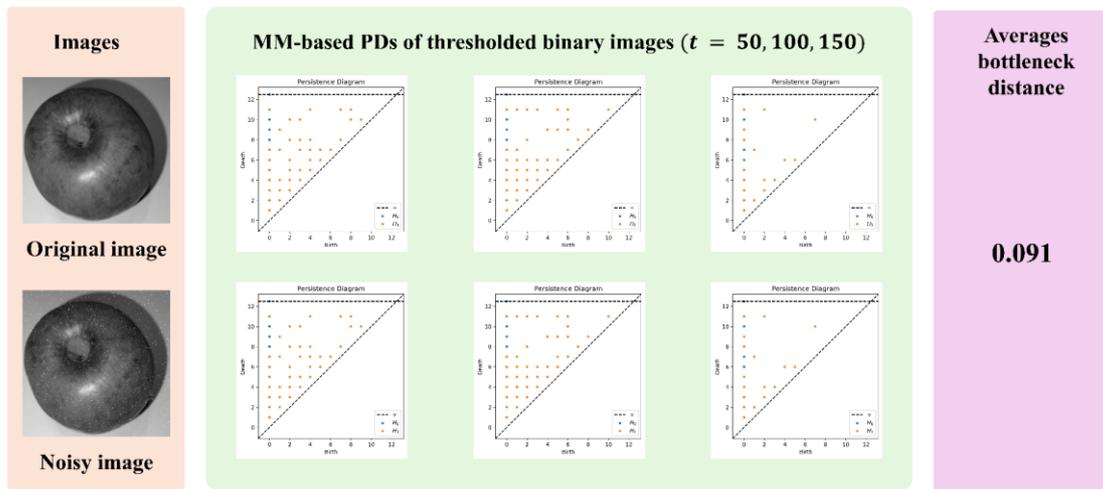

Figure 3. Illustration of a noisy grayscale image $f$ with pixel values ranging from 0 to 255, the thresholded image $g$ obtained using threshold 150, and the images produced by applying erosion $\epsilon_B$, dilation $\delta_B$, opening $O_B$, and closing $C_B$ with the $3 \times 3$ square $B = S_3$ as the structuring element (see Equation (4)).

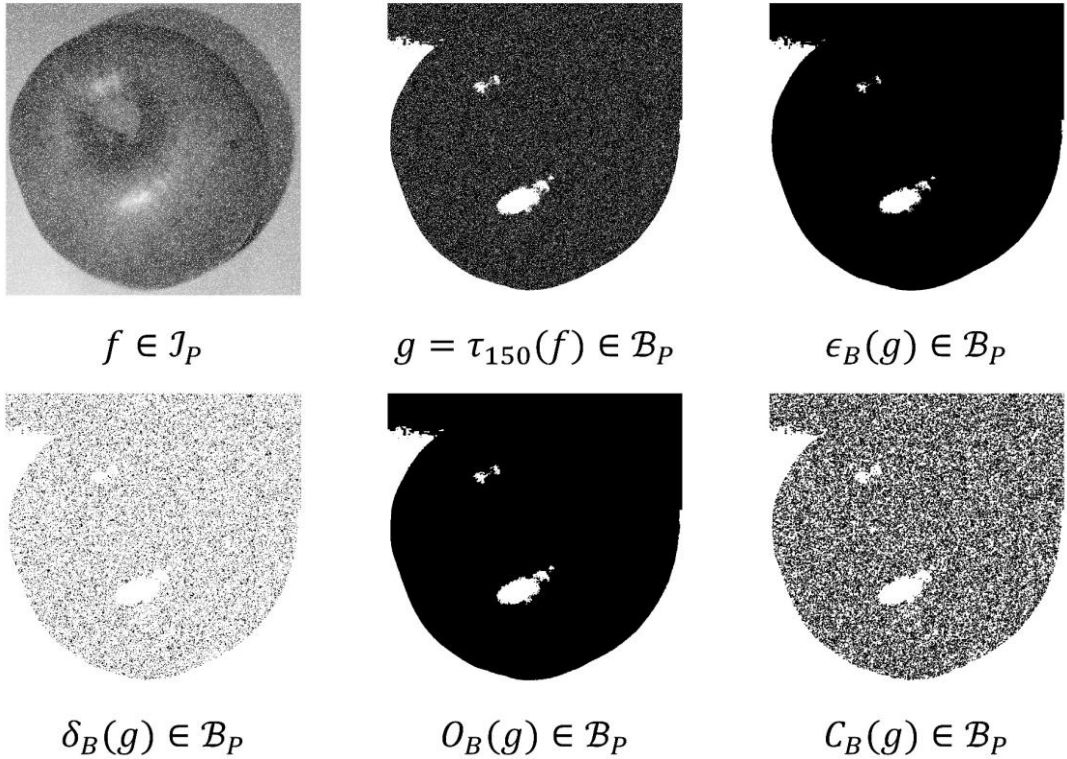

Figure 4. Illustrative examples of an image domain, a grayscale image, and a sequence of binary images with the associated filtrations of topological spaces. (a) A 5 × 5 rectangular image domain $P \subseteq \mathbb{Z}^2$. (b) A grayscale image $f$ defined on $P$ with pixel values in the range {1,2,3,4}. (c) An alternative visualization of the grayscale image $f$, where each pixel is represented by a gray square whose intensity reflects the pixel value; darker squares correspond to lower pixel values. (d)–(g) A sequence of binary images $f_1 \geq f_2 \geq f_3 \geq f_4$, where each $f_i$ is obtained by thresholding $f$ via $\tau_i$. (h)–(k) The filtration of topological spaces corresponding to the sequence of binary images shown in (d)–(g).

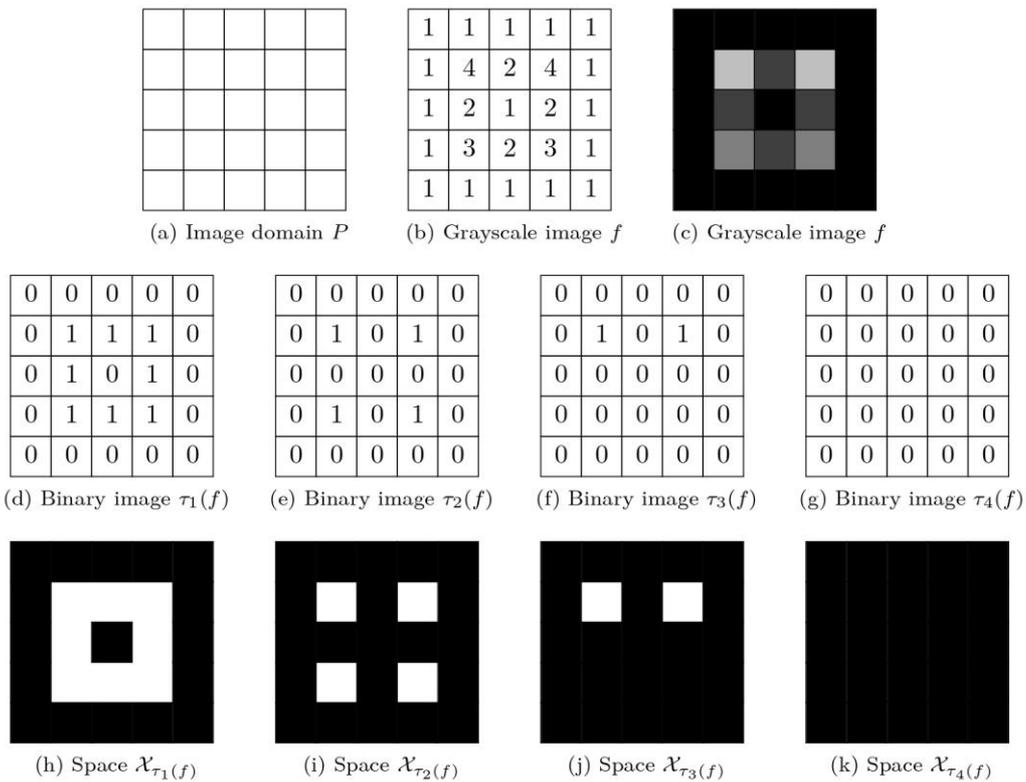

Figure 5. An example of a filtration shared with Figure 4, together with the corresponding persistence diagram and persistence barcode. In the first row (panels (a)–(d)), each topological space is annotated with its Betti pair $(\beta_0, \beta_1)$, indicating the numbers of connected components and loop structures, respectively.

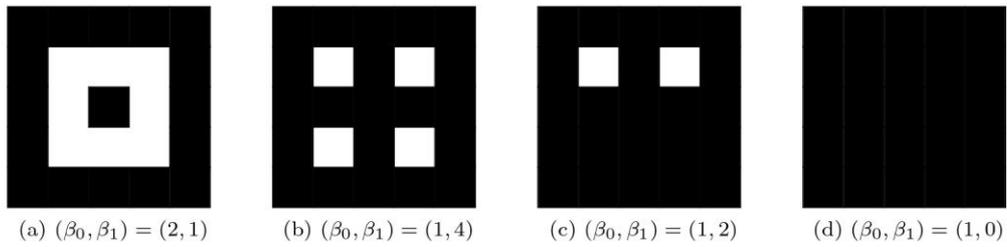

(a) $(\beta_0, \beta_1) = (2, 1)$  (b) $(\beta_0, \beta_1) = (1, 4)$  (c) $(\beta_0, \beta_1) = (1, 2)$  (d) $(\beta_0, \beta_1) = (1, 0)$

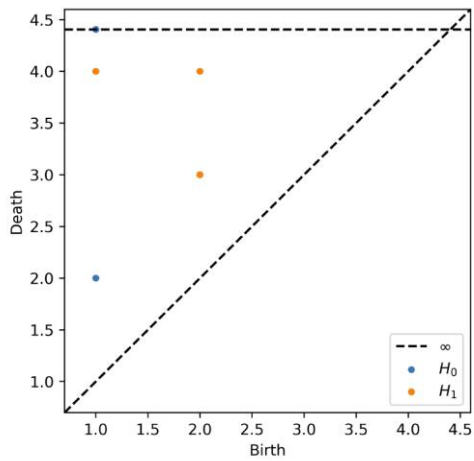
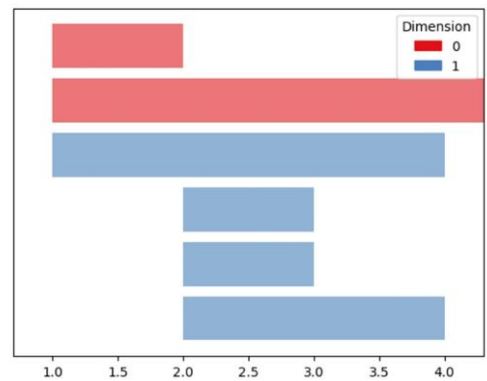

(e) Persistence diagram        (f) Persistence barcode

Figure 6. An illustrative example of a binary image, together with the induced 0th and 1st persistence diagrams of the opening filtration with squared SEs $S_n$ defined in Equation (4) and the histogram that counts the death values of PIs, is shown. The six hole structures in the image and their corresponding 1st PIs are annotated by indices 1 through 6.

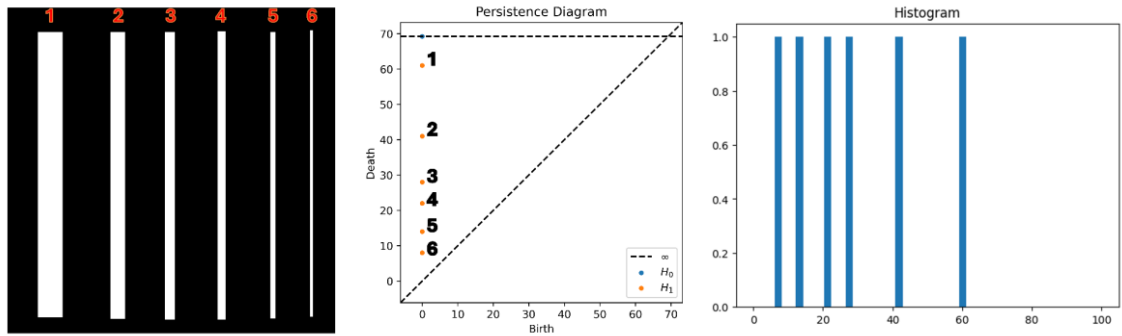

Figure 7. An example of erosion- and opening-based filtrations constructed from a binary image, along with their corresponding persistence diagrams. The histograms depict the distributions of death values of PIs in the 1st persistence diagrams ($PD_1$) induced by the two filtrations.

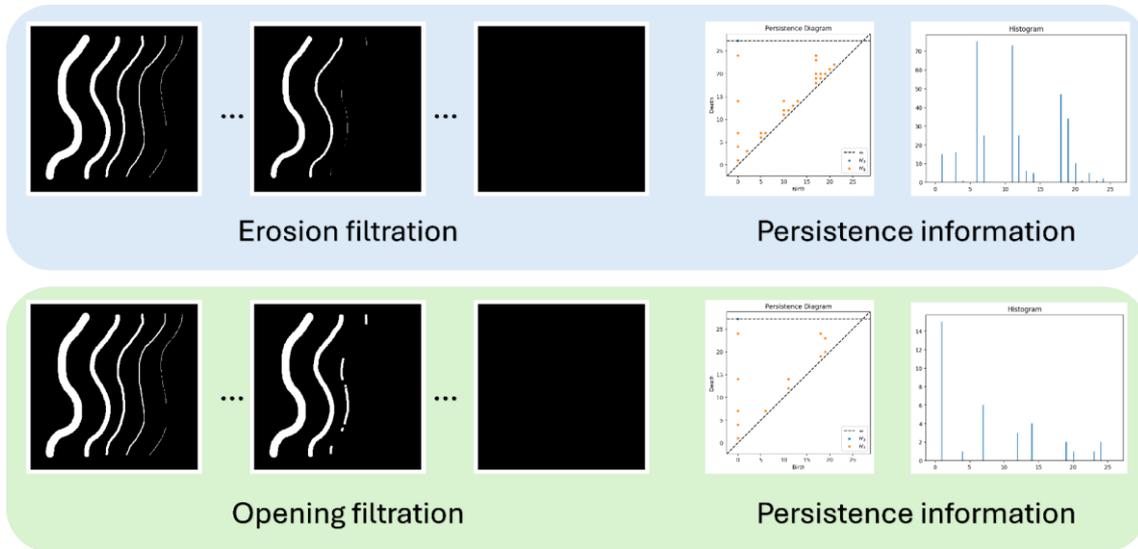

Figure 8. Examples of binary images with their PDs and PBs induced by dilation-based filtrations.

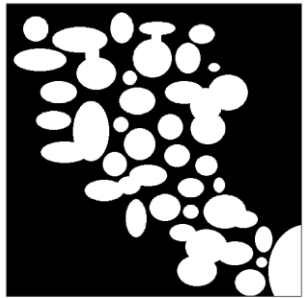 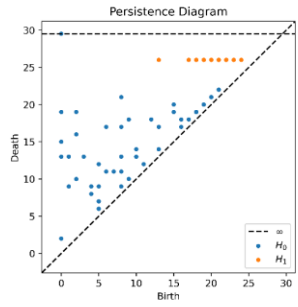 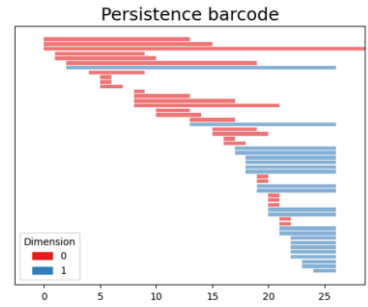

Input image — Persistence diagrams — Persistence barcode

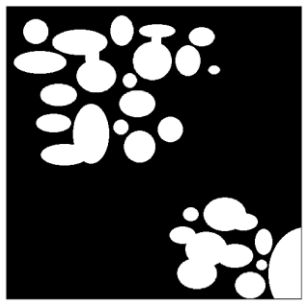 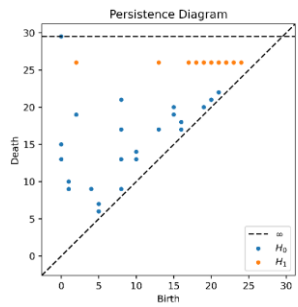 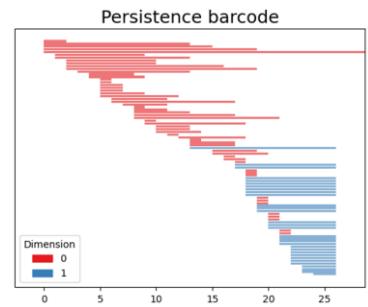

Input image — Persistence diagrams — Persistence barcode

Figure 9. A binary image together with its 0th and 1st PDs induced by dilation-based filtrations. The image contains two one-dimensional holes of identical shape and size, with the corresponding PIs highlighted by annotations.

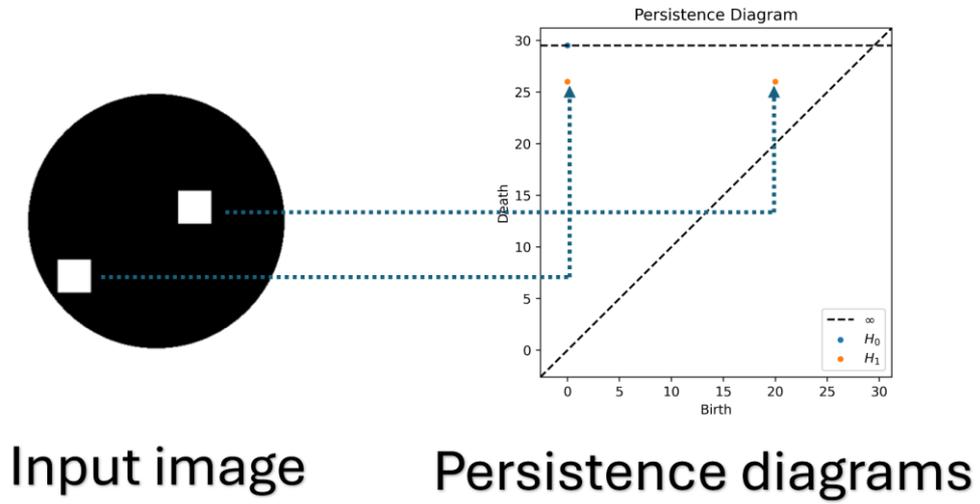

Figure 10. The computational pipeline of MMPersistence for grayscale images, based on the illustration in Figures 1 and 2.

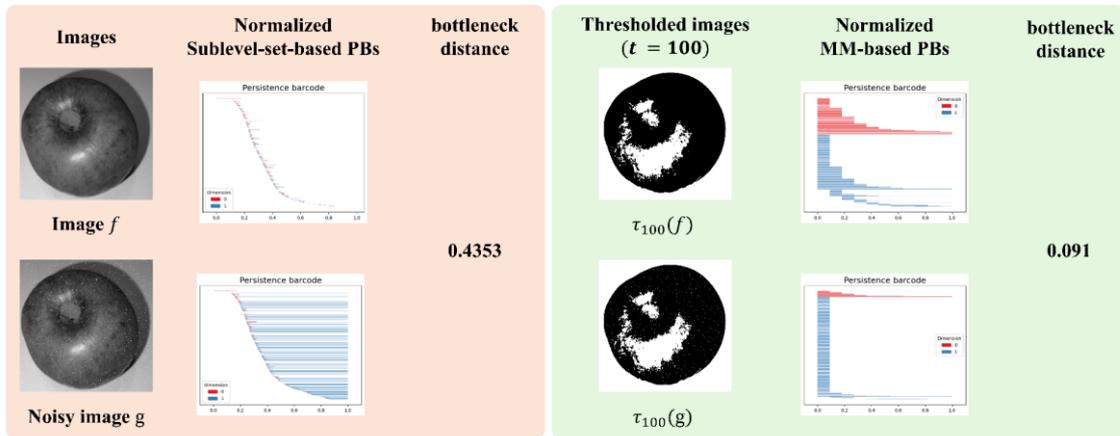